# From Linear Risk to Emergent Harm: Complexity as the Missing Core of AI Governance

Author

Hugo Roger Paz

Faculty of Exact Sciences and Technology,

National University of Tucumán (UNT)

Argentina

ORCID: 0000-0003-1237-7983

Contact: hpaz@herrera.unt.edu.ar

## ABSTRACT

Risk-based AI regulation has become the dominant paradigm in AI governance, promising proportional controls aligned with anticipated harms. This paper argues that such frameworks often fail for structural reasons: they implicitly assume linear causality, stable system boundaries, and largely predictable responses to regulation. In practice, AI operates within complex adaptive socio-technical systems in which harm is frequently emergent, delayed, redistributed, and amplified through feedback loops and strategic adaptation by system actors. As a result, compliance can increase while harm is displaced or concealed rather than eliminated.

We propose a complexity-based framework for AI governance that treats regulation as intervention rather than control, prioritises dynamic system mapping over static classifications, and integrates causal reasoning and simulation for policy design under uncertainty. The aim is not to eliminate uncertainty, but to enable robust system stewardship through monitoring, learning, and iterative revision of governance interventions.

**VERSION:** V1.0.

# 1. INTRODUCTION

Over the past few years, artificial intelligence has moved from a technical concern to a central object of public policy. Governments, regulators, and international organisations now routinely frame AI as a source of risk—to safety, fairness, privacy, employment, and democratic processes. In response, a growing body of regulation has adopted risk-based approaches (OECD, 2019), seeking to classify AI systems, anticipate potential harms, and impose controls proportional to their presumed impact.

Despite their apparent sophistication, these regulatory efforts share a common and rarely questioned assumption: that the harms associated with AI can be identified, enumerated, and mitigated through essentially linear reasoning. Risks are treated as attributes of systems, impacts as relatively localised outcomes, and regulation as an external corrective applied to an otherwise stable technological object.

## 1.1. This paper argues that this assumption is fundamentally flawed.

AI systems do not operate as isolated artefacts but as components of complex adaptive socio-technical systems (Holland, 1992). Their behaviour, impacts, and risks emerge from interactions between algorithms, institutions, markets, users, incentives, and regulatory constraints themselves. In such systems, harm is not merely a direct consequence of design choices, nor a predictable extension of technical capability. Rather, it is often emergent, delayed, redistributed, and amplified through feedback loops (Sterman, 2000; Meadows, 2008) that evade linear models of control.

As a result, many well-intentioned AI governance frameworks risk failing not because they underestimate technological power, but because they misunderstand the nature of the system they are attempting to govern.

## 1.2. The Problem With Linear Risk Thinking

Risk-based regulation has intuitive appeal. It promises proportionality, accountability, and foresight. By classifying AI systems according to predefined risk categories—"low", "high", or "unacceptable"—policymakers seek to match regulatory intensity to anticipated harm. This approach implicitly assumes that risks can be specified ex ante, that causal pathways are sufficiently stable, and that regulatory interventions will produce largely predictable effects.

These assumptions hold reasonably well for simple or merely complicated systems, where cause–effect relationships are stable and interventions scale linearly. They break down, however, when applied to complex adaptive systems.

In complex systems, interventions do not simply reduce or eliminate risk; they reconfigure the system itself. Actors adapt to constraints, incentives shift, behaviours reorganise, and new equilibria emerge. Regulatory rules become part of the system's environment, shaping strategies in ways that often generate second- and third-order effects far removed from the original policy intent.

In such contexts, the most consequential harms are frequently not those explicitly targeted by regulation, but those that emerge indirectly through adaptation, displacement, and unintended coupling across system levels.

## 1.3. AI as a complex adaptive system

Treating AI as a complex adaptive system implies several departures from dominant governance paradigms.

First, non-linearity: small regulatory changes can produce disproportionate effects, while large interventions may have surprisingly limited impact. The relationship between control and outcome is not monotonic.

Second, emergence: harms are not always traceable to individual components or decisions. They arise from interactions—between models and markets, compliance requirements and business strategies, technical safeguards and organisational incentives.

Third, feedback and adaptation: regulated actors do not passively absorb constraints. They learn, optimise, and reorganise, often in ways that redistribute risk rather than eliminate it. Over time, the system co-evolves with its regulatory environment.

Fourth, temporal delay and path dependence: the most significant effects of regulation may appear long after implementation, shaped by historical contingencies and irreversible trajectories that cannot be undone by incremental adjustment.

These properties are not peripheral complications. They are defining characteristics of the system itself. Ignoring them does not simplify governance; it merely obscures the mechanisms through which harm actually arises.

## 1.4. From risk control to system design

If AI governance is to be effective, it must shift from attempting to control risks to designing interventions within complex systems. This requires a different regulatory mindset—one that treats policies not as final solutions, but as perturbations whose effects must be explored, tested, and revised.

Rather than asking "How risky is this AI system?", the more appropriate question becomes: "How does this intervention reshape the dynamics of the system, and what new patterns of behaviour might it generate?"

This paper proposes a complexity-based framework for AI governance that integrates causal reasoning, simulation, and adaptive evaluation. The aim is not to eliminate uncertainty—an impossible task in complex systems—but to make uncertainty explicit, navigable, and institutionally manageable.

By foregrounding complexity as the core analytical lens, AI governance can move beyond symbolic compliance and towards genuinely informed decision-making—one capable of anticipating emergent harm, learning from policy experiments, and adapting to the evolving socio-technical landscape.

## 2. WHAT COMPLEXITY REALLY MEANS (AND WHY IT MATTERS FOR AI GOVERNANCE)

The term complexity is frequently invoked in discussions of artificial intelligence and regulation, yet it is rarely taken seriously as an analytical commitment. More often than not, complexity is used as a rhetorical label—a synonym for difficulty, scale, or uncertainty—rather than as a description of a system governed by qualitatively different principles than those assumed by linear policy design.

This section clarifies what is meant by complexity in a strict sense, and why failing to recognise its implications leads to systematic regulatory blind spots.

### 2.1 Complexity is not complication (Mitchell, 2009)

A complicated system may contain many components, intricate internal structures, and sophisticated technical mechanisms, yet remain fundamentally predictable. Given sufficient information, expertise, and computational power, its behaviour can be decomposed, analysed, and controlled. Most traditional engineering systems fall into this category.

A complex system is different in kind, not degree.

Complex systems are characterised by interactions rather than components, patterns rather than parts, and behaviours that cannot be inferred by analysing elements in isolation. Crucially, complexity does not arise from scale alone, but from non-linear coupling between heterogeneous actors operating under adaptive incentives.

Artificial intelligence systems deployed in real-world contexts are not merely complicated technologies. They are embedded within economic, organisational, legal, and cultural environments that respond to both technological capability and regulatory constraint. Once deployed, they participate in feedback-rich ecosystems where control is partial, provisional, and often illusory.

Treating such systems as merely complicated invites a false sense of governability.

### 2.2 Core properties of complex adaptive systems

Several properties of complex systems are particularly relevant for AI governance.

**Non-linearity**
In complex systems, effects do not scale proportionally with causes. Small regulatory changes may trigger large behavioural shifts, while extensive compliance efforts may yield marginal impact. This undermines policy designs based on incremental tuning or proportional control.

**Emergence**
System-level outcomes arise from interactions between agents rather than from the intentions or properties of individual components. In AI contexts, harms such as discrimination, exclusion, or market distortion often emerge from the interaction between models, data practices, institutional incentives, and user behaviour—none of which may be problematic in isolation.

**Adaptation and learning**
Actors within the system—developers, firms, institutions, users—adapt strategically to regulatory constraints. Over time, this produces co-evolution between technology and governance, where regulation reshapes behaviour, and behaviour reshapes the effective meaning of regulation.

**Feedback loops**
Regulatory interventions feed back into the system, altering incentives and information flows. These feedbacks may reinforce initial dynamics, dampen them, or generate entirely new trajectories. Importantly, feedback effects are often delayed and indirect, complicating attribution and evaluation.

**Path dependence and irreversibility**
Once certain trajectories are established—through market dominance, infrastructural lock-in, or institutional routines—they become difficult or impossible to reverse. Regulatory timing therefore matters as much as regulatory content.

These properties are not theoretical abstractions. They describe the operational reality of AI systems deployed at scale.

## 2.3 Why dominant governance models fail under complexity

Most contemporary AI governance frameworks implicitly rely on a linear model of causality. Risks are identified, harms are anticipated, and regulatory controls are imposed to prevent or mitigate undesirable outcomes. This logic presumes that risks are stable, that interventions act locally, and that system behaviour remains largely unchanged by the act of regulation itself.

Under conditions of complexity, these presumptions fail.

First, risks are context-dependent and endogenous. They evolve as actors respond to both technology and policy. A system deemed "low-risk" under one configuration may become high-risk under another, not due to technical change, but due to shifts in deployment incentives or compliance strategies.

Second, regulation often displaces rather than eliminates harm. Constraints imposed at one level of the system may push risk downstream, upstream, or laterally—into less visible domains where monitoring is weaker and accountability diffused.

Third, compliance becomes performative. Actors learn to optimise for regulatory categories rather than substantive outcomes, producing systems that satisfy formal requirements while preserving or even amplifying underlying harms.

In such environments, governance that ignores complexity does not merely underperform—it actively generates blind spots.

**2.4 Complexity as a design constraint, not a slogan**

Acknowledging complexity does not mean abandoning governance, nor does it imply regulatory paralysis. On the contrary, it imposes a more demanding standard on policy design.

Complexity requires regulators to treat policies as interventions in evolving systems, not as static rules applied to stable objects. This entails:

- anticipating adaptive responses rather than assuming compliance,
- evaluating system-level dynamics rather than isolated outcomes,
- accepting uncertainty as structural rather than residual,
- and designing institutions capable of learning over time.

In this view, the central failure of current AI governance is not insufficient enforcement or inadequate ethical framing, but a misalignment between the nature of the system and the logic of its regulation.

Until complexity is treated as the analytical core of AI governance—rather than as a background complication—efforts to manage risk will remain reactive, fragmented, and prone to emergent harm.

## 3. THE FAILURE OF RISK-BASED AI REGULATION

Risk-based regulation has become the dominant paradigm (NIST, 2023) in contemporary AI governance. From national frameworks to supranational initiatives, policymakers increasingly seek to classify AI systems according to predefined levels of risk and to calibrate regulatory obligations accordingly. At first glance, this approach appears pragmatic, flexible, and aligned with long-standing regulatory traditions in domains such as finance, medicine, and environmental protection.

Yet when examined through the lens of complexity, the risk-based model reveals a series of structural weaknesses that are not incidental flaws, but consequences of its underlying assumptions.

**3.1 The implicit linearity of risk-based frameworks**

Risk-based regulation rests on a deceptively simple logic:

identify potential harms, estimate their likelihood and severity, and apply controls proportional to the assessed risk.

This logic presumes a world in which:

- ➢ risks can be meaningfully enumerated ex ante,
- ➢ causal pathways are sufficiently stable to support prediction, and
- ➢ regulatory interventions act as external constraints that do not fundamentally alter system behaviour.

Such assumptions are defensible in domains where systems are relatively closed, actors are tightly regulated, and feedback effects are limited. They become untenable, however, when applied to AI systems embedded in open, adaptive, and incentive-driven socio-technical environments.

In complex systems, risk is not a static property of artefacts. It is a dynamic outcome of interaction.

**3.2 Risk classification in a moving system**

Risk-based AI governance typically relies on categorical distinctions: high-risk versus low-risk (European Union, 2024) applications, acceptable versus unacceptable uses, general-purpose versus domain-specific systems. These classifications assume that meaningful boundaries can be drawn around systems and that their behaviour remains largely invariant across contexts.

In practice, AI systems are reconfigured, repurposed, recombined, and redeployed in ways that routinely undermine such classifications. A model deemed low-risk in isolation may generate high-risk outcomes once integrated into a broader decision pipeline. Conversely, systems classified as high-risk may be fragmented, modularised, or rebranded to avoid regulatory thresholds without materially reducing harm.

Under complexity, classification does not merely describe reality—it reshapes it. Actors learn to design systems around regulatory categories, producing artefacts that are optimised for compliance rather than for systemic safety.

**3.3 Adaptive behaviour and regulatory gaming**

One of the most persistent blind spots of risk-based regulation is its treatment of regulated actors as largely passive recipients of rules. In reality, developers, firms, and institutions respond strategically to regulatory environments.

As regulatory obligations increase with assessed risk, incentives emerge to:

- ➢ minimise formal exposure rather than substantive impact,

- shift functionality across organisational or jurisdictional boundaries,
- externalise risk to downstream users or intermediaries, and
- substitute regulated practices with adjacent, less visible alternatives.

These responses are not pathological; they are rational adaptations within competitive and resource-constrained environments. From a complexity perspective, they represent normal system behaviour under constraint.

The result is a familiar pattern: apparent regulatory success at the surface level, accompanied by the re-emergence of harm in altered forms elsewhere in the system.

**3.4 The displacement, not elimination, of harm**

Perhaps the most consequential failure of risk-based AI governance is its tendency to displace harm rather than eliminate it.

When regulation targets specific applications, sectors, or technical features, pressure propagates through the system. Risk migrates:

- from regulated to unregulated domains,
- from visible to opaque actors,
- from centralised to decentralised infrastructures,
- or from technical to organisational failure modes.

Because these shifts often occur across institutional and temporal boundaries, they escape conventional monitoring and evaluation mechanisms. Harm appears reduced in the regulated domain, while increasing elsewhere—sometimes with greater intensity and less accountability.

From a linear perspective, this looks like partial success. From a systems perspective, it is a predictable failure mode.

**3.5 Compliance without control**

As risk-based regimes mature, compliance itself becomes a dominant activity. Documentation, audits, certifications, and procedural safeguards proliferate. While such mechanisms can improve transparency and accountability, they also create a false sense of control.

In complex systems, satisfying formal requirements does not guarantee desired outcomes. Indeed, high compliance capacity may correlate with greater ability to manage appearances while preserving underlying dynamics unchanged.

When governance focuses on risk scores, checklists, and static classifications, it risks mistaking regulatory artefacts for systemic understanding. Control is inferred from process rather than demonstrated through effect.

### 3.6 Why better risk models are not enough

A common response to these critiques is to propose more sophisticated risk assessment (ISO/IEC, 2023): finer categories, better metrics, more data, more expert oversight. While such improvements may address surface-level deficiencies, they do not resolve the core issue.

The failure of risk-based AI governance is not primarily epistemic. It is ontological.

No amount of refinement can transform a linear control paradigm into an effective tool for governing complex adaptive systems. As long as governance treats risk as an attribute to be managed rather than as an emergent property of system dynamics, it will remain vulnerable to unintended consequences.

### 3.7 Toward a different question

The central question for AI governance should therefore shift.

Instead of asking:
"How risky is this AI system?"

We should ask:
"How does this regulatory intervention alter the behaviour of the system, and what new dynamics does it set in motion?"

This shift does not reject risk assessment; it re-embeds it within a broader understanding of system design, adaptation, and emergence. Only by doing so can governance move from symbolic control to meaningful influence.

## 4. A COMPLEXITY-BASED FRAMEWORK FOR AI GOVERNANCE

If risk-based regulation fails because it assumes linear control over non-linear systems, the solution is not to abandon governance, but to redesign it around a more accurate model of reality. Governing AI as a complex adaptive system requires a shift in both analytical focus and institutional practice.

This section outlines a complexity-based framework for AI governance, grounded in three core principles: intervention over control, dynamics over categories, and learning over static compliance.

### 4.1 Regulation as intervention, not control

In complex systems, governance cannot function as external command-and-control. Any regulatory action becomes part of the system it seeks to influence, altering incentives, behaviours, and information flows.

A complexity-based framework therefore treats regulation as a deliberate perturbation rather than as a final solution. Policies are understood as interventions that reshape system dynamics, often in unpredictable ways, and whose effects must be continuously observed and reassessed.

This perspective reframes regulatory success. The goal is not to "eliminate risk", but to steer system behaviour away from harmful attractors and towards more resilient configurations.

**4.2 From static classification to dynamic system mapping**

Rather than classifying AI systems into fixed risk categories, a complexity-oriented approach focuses on mapping interactions (Meadows, 2008):

- Which actors are involved?
- How are decisions distributed across technical, organisational, and institutional layers?
- Where do incentives align or conflict?
- Which feedback loops amplify or dampen harm?

This mapping does not seek exhaustive prediction. Its purpose is to identify structural leverage points—locations where small, well-designed interventions can produce disproportionate effects.

Importantly, this shifts attention away from individual models and toward decision pipelines, organisational practices, and systemic coupling between technology and policy.

**4.3 Causal reasoning beyond attribution**

Complexity does not negate causality; it complicates it.

In AI governance, causal reasoning must move beyond simple attribution ("this model caused that harm") toward causal configurations that capture how multiple factors interact over time. Structural causal models provide a disciplined way to articulate assumptions, test counterfactuals, and distinguish between correlation and intervention-relevant causation.

Within a complexity-based framework, causal analysis serves three purposes:

1. to clarify which mechanisms plausibly generate observed harms,

2. to explore how alternative interventions might alter system trajectories, and

3. to avoid policy decisions based solely on surface-level associations.

Causality, in this sense, becomes a tool for design under uncertainty, not for retrospective blame.

**4.4 Simulation as a governance instrument**

One of the most underutilised tools in AI governance is simulation. In complex systems, simulation offers a way to explore possible futures without incurring real-world cost.

Agent-based models, system dynamics simulations, and scenario-based causal experiments can be used to test regulatory ideas before deployment. While such models are necessarily simplified, their value lies not in prediction, but in revealing unintended consequences, adaptive responses, and emergent patterns.

Simulation transforms governance from a purely normative exercise into an exploratory one. Policies become hypotheses to be tested, revised, and iterated, rather than static rules imposed under conditions of false certainty.

**4.5 Adaptive governance and institutional learning**

Complexity demands institutions capable of learning.

Static regulatory regimes assume that rules, once enacted, remain valid until formally revised. In fast-evolving socio-technical systems (Jasanoff, 2004), this assumption is untenable. A complexity-based framework therefore embeds continuous monitoring, feedback, and revision into governance itself.

This implies:

➢ treating regulatory deployment as provisional,

➢ designing evaluation mechanisms focused on system-level outcomes rather than procedural compliance, and

➢ granting institutions the authority to adapt rules in response to observed dynamics.

Such adaptability does not weaken governance; it strengthens it by aligning institutional behaviour with the reality of system evolution.

**4.6 Governing uncertainty, not denying it**

Perhaps the most profound implication of a complexity-based approach is its treatment of uncertainty. Rather than framing uncertainty as a temporary knowledge gap to be closed, it recognises uncertainty as a structural feature (Cilliers, 1998) of complex systems.

Effective AI governance does not promise certainty or control. It promises robustness, resilience, and the capacity to respond when surprises occur.

This requires a cultural shift within regulatory institutions: away from the illusion of comprehensive foresight and toward a practice of responsible experimentation, transparency about limitations, and humility in the face of emergent behaviour.

### 4.7 From compliance to system stewardship

Taken together, these elements redefine the role of AI governance. Regulators are no longer mere enforcers of compliance, nor arbiters of abstract risk scores. They become stewards of evolving socio-technical systems, responsible for shaping conditions under which AI can develop without generating disproportionate harm.

This role is more demanding than traditional regulation. It requires interdisciplinary competence, methodological sophistication, and institutional courage. But it is also the only role consistent with the nature of the system being governed.

## 5. WHEN LINEAR REGULATION PRODUCES EMERGENT HARM: AN ILLUSTRATIVE FAILURE

Abstract arguments about complexity often fail to persuade not because they are incorrect, but because they remain detached from concrete policy experience. This section presents an illustrative scenario—not as a case study tied to a specific jurisdiction, but as a structurally realistic pattern that recurs across AI governance contexts.

The purpose is not to single out regulatory failure, but to expose a mechanism: how linear regulatory logic, when applied to complex adaptive systems, can systematically generate emergent harm.

### 5.1 The regulatory intent

Consider a regulator implementing a risk-based AI framework aimed at preventing discrimination in automated decision-making. Systems used in high-stakes domains—such as hiring, credit scoring, or access to public services—are classified as "high-risk" and subjected to strict requirements: explainability, bias audits, documentation, and human oversight.

The regulatory intent is clear and defensible. By increasing scrutiny where harm is most likely, the regulator seeks to reduce discriminatory outcomes and protect vulnerable populations.

At the level of intent, the policy is sound.

### 5.2 The adaptive response

Once the regulation is enacted, regulated actors begin to adapt. Firms restructure decision pipelines to reduce formal exposure. Instead of a single high-risk AI system making consequential decisions, functionality is distributed across multiple components: pre-filtering models, recommendation systems, human-in-the-loop processes, and post-hoc review layers. Each component, considered in isolation, falls below regulatory thresholds.

From a compliance perspective, this appears as success. The organisation demonstrates adherence to the framework while maintaining operational efficiency.

From a systems perspective, something more subtle occurs.

**5.3 The emergence of a new harm configuration**

The fragmentation of decision-making redistributes responsibility and obscures accountability. Bias audits focus on individual components, while discriminatory effects arise from their interaction. Human oversight becomes nominal, as operators rely on upstream signals shaped by algorithmic filtering. Errors propagate through the system without triggering regulatory alarms, because no single element violates formal requirements.

Meanwhile, smaller actors—lacking the resources to implement complex compliance architectures—exit the market or avoid regulated domains altogether. Market concentration increases, reinforcing the position of large firms best equipped to navigate regulatory complexity.

The original harm—discrimination—has not disappeared. It has shifted form, moved location, and become harder to detect.

This is not regulatory failure by accident. It is emergent harm by design, produced by the interaction between linear regulation and adaptive system behaviour.

**5.4 Why the failure remains invisible**

From a traditional evaluation standpoint, the regulation appears effective. Formal compliance rates are high. Documentation is complete. Audits report adherence to standards.

The system passes its own tests.

What remains invisible are the second- and third-order effects: the redistribution of risk, the erosion of meaningful oversight, and the consolidation of power in actors capable of absorbing regulatory friction.

Because these effects unfold gradually and across institutional boundaries, they escape conventional metrics of regulatory success. The framework measures what it was designed to measure—and misses what it was not designed to see.

**5.5 The complexity diagnosis**

From a complexity perspective, the outcome is unsurprising. The regulation acted as a perturbation in an adaptive system. Actors responded rationally to incentives. Feedback loops reinforced compliant but strategically optimised behaviour. Over time, the system settled into a new configuration—one that satisfied formal requirements while reproducing underlying harm.

Crucially, no individual decision caused the failure. The harm emerged from system-level dynamics, not from malicious intent or technical malfunction.

This is precisely the kind of failure that linear risk models are structurally incapable of anticipating.

**5.6 What a complexity-aware approach would change**

A governance framework grounded in complexity would not seek to prevent adaptation; it would anticipate it.

Rather than focusing exclusively on component-level compliance, it would examine interaction effects across decision pipelines. Rather than assuming that harm reduction follows from formal oversight, it would monitor system-level outcomes over time. Rather than treating regulation as a static rule set, it would treat it as an ongoing experiment, subject to revision in light of observed dynamics.

Most importantly, it would recognise that the absence of visible violations does not imply the absence of harm.

**5.7 The lesson**

This scenario is not hypothetical in the sense of speculative fiction. It reflects a pattern repeatedly observed whenever linear governance models confront complex adaptive systems.

The lesson is not that AI regulation is futile, nor that risk-based approaches are useless in all contexts. It is that without a complexity-aware design logic, regulation will continue to chase symptoms while reproducing causes.

Effective AI governance must therefore move beyond the question of how to control risk, and confront the more difficult task of how to shape system dynamics under conditions of uncertainty and adaptation.

## 6. IMPLICATIONS FOR POLICY, INSTITUTIONS, AND RESEARCH

Reframing AI governance around complexity is not a semantic exercise, nor an academic refinement. It carries concrete implications for how policies are designed, how institutions operate, and how research informs decision-making. This section outlines the practical consequences of taking complexity seriously—and the risks of continuing to ignore it.

**6.1 Implications for policy design**

A complexity-based approach fundamentally alters what it means to design AI policy.

First, policies should be conceived as provisional interventions, not as definitive solutions. This implies embedding mechanisms for revision, rollback, and recalibration from the outset, rather than treating policy change as an exceptional or politically costly event.

Second, regulatory focus must shift from static system properties to dynamic system behaviour. Instead of asking whether an AI system meets predefined criteria at a single point in time, governance should prioritise how systems evolve, interact, and adapt under regulatory pressure.

Third, policy evaluation must extend beyond compliance metrics. Indicators of success should include system-level outcomes: shifts in market structure, changes in decision-making practices, redistribution of risk, and the emergence of new failure modes. Without such indicators, governance remains blind to its own effects.

Finally, complexity implies that policy sequencing matters. Early interventions can lock systems into trajectories that later regulation struggles to correct. This places a premium on anticipatory design and cautious deployment, particularly in foundational infrastructures.

**6.2 Implications for regulatory institutions**

Governing complex adaptive systems requires institutions with capabilities that extend beyond legal enforcement.

Regulatory bodies must develop system literacy: the ability to reason about feedback, adaptation, and emergence across socio-technical domains. This does not mean that regulators must become technologists, but that they must be equipped to interrogate models, incentives, and organisational dynamics at a structural level.

Institutional design must also support learning over time. This includes access to longitudinal data, the capacity to run simulations or scenario analyses, and organisational cultures that reward revision rather than penalise it. In complex systems, persistence in ineffective rules is not stability; it is fragility.

Crucially, complexity-aware governance requires institutional humility. Regulators must be willing to acknowledge uncertainty, communicate limitations, and adapt in response to unexpected outcomes. Such humility is not a weakness, but a prerequisite for credibility in environments where certainty is structurally unattainable.

**6.3 Implications for research and expertise**

The dominance of risk-based governance has shaped research agendas accordingly, privileging classification, benchmarking, and post-hoc impact assessment. A shift toward complexity demands a parallel shift in the production of knowledge.

Research that informs AI governance must engage with system dynamics, not merely system performance. This includes studying adaptive behaviour, regulatory gaming, organisational response, and long-term structural effects.

Methodologically, this favours approaches capable of exploring interaction and emergence: causal modelling, agent-based simulation, system dynamics, and mixed qualitative–quantitative analysis. No single method is sufficient; what matters is coherence with the nature of the system under study.

Expertise, in this context, is not defined by proximity to technology alone, but by the ability to reason across levels—from technical mechanisms to institutional incentives and societal outcomes. The most valuable contributions to AI governance may therefore come from interdisciplinary spaces that remain marginal in current policy debates.

### 6.4 The cost of ignoring complexity

Failing to incorporate complexity into AI governance does not preserve the status quo; it actively increases systemic risk.

Linear regulatory frameworks create the illusion of control while leaving underlying dynamics unchecked. Over time, this gap between appearance and reality widens, eroding trust in institutions and amplifying the very harms governance seeks to prevent.

In complex systems, delayed recognition of failure is often more damaging than early acknowledgement of uncertainty. By the time emergent harms become visible, path dependence and institutional inertia may render meaningful correction impossible.

Ignoring complexity is therefore not a neutral stance. It is a decision to govern blind.

### 6.5 Conclusion: from control to stewardship

This paper has argued that the central challenge of AI governance is not technological unpredictability, ethical disagreement, or insufficient data. It is a misalignment between the complexity of the system and the simplicity of the models used to govern it.

Risk-based approaches, while useful in limited contexts, are structurally incapable of managing emergent harm in complex adaptive systems. Effective governance must instead embrace complexity as its analytical core, treating policies as interventions, institutions as learning systems, and uncertainty as an irreducible condition.

The task of AI governance is not to control technology, but to steward evolving socio-technical systems under conditions of adaptation and surprise. Whether current institutions are willing to assume this role remains an open question. What is clear, however, is that without such a shift, governance will continue to lag behind the dynamics it seeks to regulate.